\documentstyle[preprint,tighten,aps,epsf]{revtex}
\begin{document}
\draft
%
%
%            A B S T R A C T
%
%
\title{
Screening of persistent currents in mesoscopic metal rings}

\author{Axel V\"{o}lker and Peter Kopietz}
\address{
Institut f\"{u}r Theoretische Physik der Universit\"{a}t G\"{o}ttingen, \\
Bunsenstra{\ss}e 9, 37073 G\"{o}ttingen, Germany}
\date{September 6, 1996}
\maketitle
\begin{abstract}
The effect of the Coulomb-interaction 
on persistent currents in disordered mesoscopic metal rings threaded by 
a magnetic flux $\phi$ is studied numerically.
We use the simplest form of {\it{self-consistent}} Hartree theory,
where the spatial variations of the 
self-consistent Hartree potential are ignored.
In this approximation the
self-consistent Hartree energies are simply obtained by
diagonalizing the non-interacting system via the Lanczos method 
and then calculating the (disorder-dependent) particle number
on the ring self-consistently. 
In the diffusive regime we find that
the variance of the total particle number
is strongly reduced, in  agreement with 
the prediction of the random-phase approximation.
On the other hand, 
the variance of the number of energy
levels in a small interval below the Fermi energy is
not affected  by the Coulomb interaction.
We argue that this implies that
the experimentally observed enhancement of the
persistent current is due to
long-range Coulomb interactions.

\end{abstract}
\pacs{PACS numbers: 73.50.Bk, 72.10.Bg, 72.15.Rn}
\narrowtext
%
%   I N T R O D U C T I O N
%
%
%
\section{Introduction}
\label{sec:Intro}

The usual diagrammatic approach\cite{Altshuler85} to disordered metals
is based on the existence of two small parameters:
the smallness of $ ( { k_F \ell })^{-1}$
(where  $k_F$ is the Fermi wave-vector and $\ell$ is the elastic mean free path)
justifies the perturbative treatment of the disorder potential, and
the smallness of the parameter $\kappa / k_F$ at high densities
(where $\kappa$ is the Thomas-Fermi screening wave-vector) 
justifies the so-called
random-phase approximation (RPA), which re-sums the most divergent
diagrams in the expansion of the Coulomb potential.
This leads to the screening of the long-range tail
of the Coulomb potential, so that the electron-electron interaction
can be taken into account via an effective short-range interaction of the Hubbard-type.

In Ref.\cite{Ambegaokar90}
this standard diagrammatic approach was used to calculate
the effect of electron-electron interactions on the average persistent
current in mesoscopic metal rings threaded by a magnetic flux
$\phi$\cite{Hund38,Buttiker83}. Surprisingly, the average  current
was found to be 
almost two orders of magnitude smaller than the experimental
result by L\'{e}vy {\it{et al.}}\cite{Levy90}.
This experiment and subsequent experiments by other groups\cite{Chandrasekhar91,Mailly93} 
have stimulated many recent theoretical works.
Nevertheless, till now a generally accepted theoretical explanation of the
experimental data\cite{Levy90,Chandrasekhar91} has not been found,
perhaps because the standard
perturbative approach is not applicable in the persistent current problem.
This has several reasons:
First of all,  the experiments\cite{Levy90,Chandrasekhar91} 
are performed at constant particle number $N$,
so that
the persistent current should be calculated from the
flux-derivative of the canonical free energy 
$F ( N, \varphi )$,
 \begin{equation}
 I ( N , \varphi ) = - \frac{e}{h} \frac{ \partial F (N , \varphi )}{\partial \varphi }
 \; \; \; ,
 \label{eq:Ican}
 \end{equation}
where $\varphi = \phi / \phi_0$ is the flux measured in
units of the flux quantum $\phi_0 = hc/e$.
On the other hand, with the  standard machinery of many-body theory 
one calculates the current $I ( \mu , \varphi )$  at constant chemical 
potential $\mu$,
which is related to the grand canonical potential $\Omega ( \mu , \varphi )$
via
 \begin{equation}
 I (\mu,  \varphi ) = - \frac{e}{h} \frac{ \partial \Omega ( \mu , \varphi )}{\partial \varphi }
 \; \; \; .
 \label{eq:Igrandcan}
 \end{equation}
Several authors\cite{Bouchiat89,Schmid91,Oppen91,Altshuler91} have noticed 
that for {\it{non-interacting}} electrons 
there exist striking differences between
the average currents
$\overline{ I ( N , \varphi ) }$ and $ \overline{ I ( \mu , \varphi)} $
(the overline denotes averaging over the disorder).
In particular, for non-interacting electrons in the diffusive regime
$\overline{I ( \mu , \varphi ) }$ is exponentially small, whereas
$\overline{ I ( N , \varphi ) }$ is not.
The question whether these qualitative differences persist
even in the presence of electron-electron
interactions has not yet been investigated.

We would like to point out two more
possible reasons for the failure 
of conventional perturbative many-body techniques 
in the persistent current problem.
The first is the existence
of two different purely geometric length scales in mesoscopic rings, namely the circumference
$L$, and the characteristic thickness $L_{\bot} $ of the rings. 
Thus, diagrams which are usually neglected because they involve
higher orders in $({k_F \ell})^{-1}$ may become important
because they are geometrically enhanced by additional
powers of $ L / L_{\bot} \gg 1$. This has first been pointed out
by B\'{e}al-Monod and Montambaux\cite{Beal92}.
Finally, it is questionable whether the conventional
RPA is valid for persistent currents.
Note that the {\it{flux-dependent}} parts of
$F ( N , \varphi )$ and $\Omega (\mu , \varphi )$ are much smaller than the 
flux-independent parts. 
The RPA has been developed to re-sum the 
leading (flux-independent) terms in the expansion of
$\Omega ( \mu , \varphi )$ in powers of the Coulomb interaction\cite{Mattuck76}.
However, it is not at all clear whether such a resummation
of a formally divergent series remains valid
for the sub-leading
flux-dependent part of $\Omega ( \mu , \varphi )$.
We shall come back to this point in Sec.\ref{sec:num}.

Recently one of us\cite{Kopietz93,Kopietz94a}
has proposed that the {\it{flux-dependent}} part of the average
Hartree energy in a mesoscopic ring is only weakly
screened, and showed that in this way the 
large magnitude of the experimentally observed 
currents\cite{Levy90} can be easily explained.
The arguments given in Refs.\cite{Kopietz93,Kopietz94a}
were simple but powerful, and made
use of the exact Hohenberg-Kohn theorem of density-functional
theory\cite{Hohenberg64}. 
Although Ref.\cite{Kopietz93} has been criticized\cite{Vignale94,Altland94},
all arguments put forward against Ref.\cite{Kopietz93}
where essentially based on the RPA.
The possibility that the RPA might not be a good approximation
for the calculation of the persistent current\cite{Kopietz94b,Kopietz94c}
was ignored.
Because even in the absence of disorder it is very difficult to
calculate the dielectric function beyond the RPA,
numerical methods seem to be the only controlled way 
to settle the issue. So far all numerical 
studies of two- and three-dimensional systems in the diffusive regime 
have found that the long-range part of the Coulomb interaction
indeed strongly enhances the average persistent 
current\cite{Berkovits95,Berkovits96,Yoshioka95,Bouzerar95}.
Among the numerical methods, the exact diagonalization
of the many-body Hamiltonian\cite{Berkovits95,Berkovits96}
is free of  any approximations. 
Unfortunately, it is very difficult to study systems
with more than approximately $10$ electrons with this method. This is certainly
too small to address the screening problem.
Calculations based on the self-consistent Hartree-Fock approximation
have been pushed to electron numbers of the order of $10^2$. Although
these calculations do not produce the exact solution of the many-body system,
they are non-perturbative by virtue of the self-consistency condition,
and are therefore more reliable than any low-order diagrammatic
calculation. 

Guided by the evidence that the enhancement 
of the persistent current is closely related
to the physics of screening, we shall
in this work examine the screening problem numerically.
As a starting point, we follow 
Ref.\cite{Altshuler91}
and use the approximate mapping
from the average canonical current
onto an effective grand canonical average,
 \begin{equation}
  \overline{I (  N, \varphi )} -
  \overline{I (  \mu^{\ast} , \varphi )} 
  \approx  \frac{- e}{  h }
  \frac{1}{2 
  \kappa ( \mu^{\ast} , \varphi )
  }
  \frac{ \partial }{\partial \varphi} 
  \Sigma^{2} ( \mu^{\ast} , \varphi )
  \; \; \; ,
  \label{eq:AGI}
  \end{equation}
where 
  $\overline{I (  \mu^{\ast} , \varphi )} $ is the average
grand canonical current at 
chemical potential $\mu^{\ast}$, and
 \begin{equation}
 \kappa ( \mu , \varphi ) =    \left(
 \frac{\partial \overline{N ( \mu , \varphi )} }{ \partial \mu } \right)_{\varphi} 
 \label{eq:kappadef}
 \end{equation}
is the average compressibility.
$\Sigma^{2} ( \mu^{\ast} , \varphi )$ is the variance
of the particle number $N ( \mu^{\ast} ,\varphi )$ in a grand canonical ensemble 
 \begin{equation}
  \Sigma^{2} ( \mu^{\ast} , \varphi )
  = \overline{N^2 ( \mu^{\ast} , \varphi )} -
   \left[ \overline{N ( \mu^{\ast} , \varphi )} \right]^2 
   \; \; \; .
 \label{eq:SigmaNdef}
 \end{equation}
The value of $\mu^{\ast}$ should be chosen such that 
the disorder- and flux-averaged particle number in the
corresponding grand canonical ensemble
agrees with the given particle number $N$ in the
original canonical ensemble, i.e.
 \begin{equation}
\int_{0}^{1} d \varphi \overline{ N ( \mu^{\ast}, \varphi )} = N
\label{eq:mustardef}
\; \; \; .
\end{equation}
Note that for non-interacting electrons in the diffusive 
regime
  $\overline{I (  \mu^{\ast} , \varphi )} $ 
is exponentially small\cite{Cheung89}.
The leading interaction contribution to
  $\overline{I (  \mu^{\ast} , \varphi )} $ 
has been calculated by Ambegaokar and Eckern\cite{Ambegaokar90}.
In this work we would like to focus on the effect of electron-electron
interactions on the 
difference between the canonical and the grand canonical current,
i.e. the term on the right-hand side of Eq.\ref{eq:AGI}. 
Recently Berkovits and Avishai\cite{Berkovits96}
studied this term via exact diagonalizations of
small systems of electrons interacting with Coulomb forces.
However, with exact diagonalizations it is impossible
to reach system sizes where the condition 
$\ell \ll L \ll \xi$ is realized. Here
$\ell$ is the elastic mean free path, $L$ is the circumference
of the ring, and $\xi$ is the localization length.
This condition defines the diffusive regime, which is
relevant for the experiments of Refs.\cite{Levy90,Chandrasekhar91}.
In this work we shall therefore take a different approach, and
evaluate the right-hand side of Eq.\ref{eq:AGI}
within the simplest possible approximation
which still contains the physics of screening, 
namely the simplified 
capacitance model\cite{Kopietz94a,Kopietz94c,Altshuler92}.
In this model the spatial dependence of the self-consistent
Hartree potential is ignored, so that it simply renormalizes
the chemical potential. 
Nevertheless, diagrammatically this model still contains
the usual infinite series of bubble diagrams which 
lead to the RPA-picture of screening. But
the model contains infinitely many other non-RPA diagrams,
so that the numerical solution of the 
self-consistent Hartree theory for this model allows us to check the accuracy of the RPA.
Note that the simplified capacitance model is easily solved numerically by
diagonalizing the {\it{non-interacting}} system with the help
of the Lanczos method, and then imposing a simple
self-consistency loop for the particle number.
The most difficult step in this procedure is the exact diagonalization,
so that in this way we can reach the same system sizes as without interactions.

\section{The simplest form of self-consistent Hartree theory}
\label{sec:cap}

To calculate the effect of electron-electron interactions
on the persistent current,
let us use the self-consistent Hartree approximation to take
the Coulomb interactions between all charges (i.e. electrons
and positively charged ions) into account.
Numerically it is convenient to use a lattice model.
The Hartree energies $E_{\alpha} ( \varphi )$ 
and wave-functions $\psi_{\alpha} ( {\bf{r}} )$ are 
obtained by solving the Schr\"{o}dinger equation
 \begin{equation}
  - \sum_{ \mu = x,y,z } 
  \left[  t_{\mu}
  \psi_{\alpha} ( {\bf{r}} + {\bf{a}}_{\mu} )
  + t_{\mu}^{\ast}
  \psi_{\alpha} ( {\bf{r}} - {\bf{a}}_{\mu} ) \right]
 + \left[
 U ( {\bf{r}} ) 
 + U_H ( {\bf{r}} ) \right] 
 \psi_{\alpha} ( {\bf{r}} ) = 
 E_{\alpha} ( \varphi )  \psi_{\alpha} ( {\bf{r}} )
 \label{eq:schroedinger}
 \; \; \; ,
 \end{equation}
where ${\bf{r}}$ labels the sites of a three-dimensional cubic lattice
with primitive vectors ${\bf{a}}_{x}$, 
${\bf{a}}_{y}$, 
${\bf{a}}_{z}$. 
To model the Aharonov-Bohm flux $\phi$ through the ring,
we call the azimuthal direction 
the $x$-direction and choose
$t_x = t e^{  \frac{ 2 \pi i \varphi}{N_x}} $,
$t_y = t_z = t$.
Here $N_x$ is the number of lattice sites
in the $x$-direction.
For convenience, from now on all energies will be measured in units of the
hopping energy $t$.
As usual, disorder is introduced via random potentials $U ( {\bf{r}} )$, which
are assumed to be independent random variables
with zero average and
uniform distribution in the interval
$[ - w/2 , w/2 ]$. The self-consistency is imposed by
requiring that the Hartree potential $U_H ( {\bf{r}} )$ satisfies
 \begin{equation}
  U_H ( {\bf{r}} )  
  = \sum_{ \stackrel{{\bf{r}}^{\prime}}{
  {\bf{r}}^{\prime} \neq {\bf{r}} }  
  }  \frac{ e^2}{ | {\bf{r}} - {\bf{r}}^{\prime} | }
  \left[  n ( {\bf{r}}^{\prime} ) - n_{+} ( {\bf{r}}^{\prime} ) \right] 
  \; \; \; ,
  \label{eq:UHdef}
  \end{equation}
with
 $n ( {\bf{r}} ) = {\sum_{\alpha}}^{\prime} | \psi_{\alpha} ( {\bf{r}} ) |^2$.
The prime indicates that
the $\alpha$-sum is in a canonical ensemble over the lowest $N$ 
solutions of Eq.\ref{eq:schroedinger}, and 
in a grand canonical ensemble over all $\alpha$ with
$E_{\alpha} < \mu$, where $\mu$ is the chemical potential.
The  number density
$n_{+} ( {\bf{r}} )$
of the positive background charge is assumed to be fixed.
Given the self-consistent solutions of Eqs.\ref{eq:schroedinger} and \ref{eq:UHdef},
the canonical persistent
current can be calculated from Eq.\ref{eq:Ican} with\cite{Hohenberg64}
 \begin{equation}
 F ( N , \varphi ) = \sum_{\alpha = 1}^{N} E_{\alpha} ( \varphi )
 - \frac{1}{2}
   \sum_{ {\bf{r}} } n ( {\bf{r}} ) U_H ( {\bf{r}} )
  \; \; \; ,
  \label{eq:FNhartree}
  \end{equation}
and the grand canonical current is obtained from Eq.\ref{eq:Igrandcan} with
 $\Omega ( \mu , \varphi ) = F ( N ( \mu , \varphi ) , \varphi ) - \mu N ( \mu , \varphi )$,
where the function $N ( \mu , \varphi )$ is now explicitly given by 
 \begin{equation}
 N (\mu , \varphi )  = \sum_{\alpha} \Theta ( \mu - E_\alpha ( \varphi ) )
 \label{eq:Nselfcon}
 \; \; \; .
 \end{equation}
In order to obtain the average current, we should solve
Eq.\ref{eq:schroedinger} self-consistently for many realizations of 
disorder and then average the result.
Of course, this is a very difficult problem,
which is beyond the scope of this work.
Instead, we shall use Eq.\ref{eq:AGI} to calculate
the difference between the canonical- and the grand canonical current approximately.

To reduce the numerical work even further without loosing the
physics of screening, 
let us replace the self-consistent Hartree potential by its 
spatial average,
$ U_H ( {\bf{r}} ) \rightarrow  \frac{1}{N_L}  \sum_{\bf{r}} U_H ( {\bf{r}} )$,
where $N_L = \sum_{\bf{r}}$ is the number of lattice sites\cite{Kopietz94a,Kopietz94c,Altshuler92}.
In this approximation $U_H$ is simply given by
 \begin{equation}
 U_H \approx \frac{e^2 (N - N_{+}) }{C_0} 
 \; \; \; , \; \; \;
 \frac{1}{C_0} =
 \frac{1}{N_L ( N_L -1 )}
 \sum_{ \stackrel{ 
 {\bf{r}} , {\bf{r}}^{\prime} }
 {{\bf{r}} \neq  {\bf{r}}^{\prime}}}  \frac{ e^2}{ | {\bf{r}} - {\bf{r}}^{\prime} | }
 \label{eq:Hselfcon}
 \; \; \; ,
 \end{equation} 
where $N = \sum_{\bf{r}} n ( {\bf{r}} )$ is the total number
of electrons on the ring, and 
$N_{+} = \sum_{\bf{r}} n_{+} ( {\bf{r}} )$ is the total number
of positively charged ions.
Note that $C_0$ is nothing but the classical capacitance of the ring.
In Fourier space the approximation \ref{eq:Hselfcon} means that 
we neglect the momentum transfer through all Coulomb lines.
Because for a {\it{thin ring}} (with $L_{\bot} \ll L$) the long-wavelength Fourier components
of the Coulomb potential depend only logarithmically on the
wave-vector, we expect that in this case the approximation
\ref{eq:Hselfcon} is sufficient for a {\it{qualitative}} estimate 
of the importance of Coulomb interactions.
Of course, in a canonical ensemble Eq.\ref{eq:Hselfcon} is a 
trivial flux-independent constant,
so that the interaction-contribution to the persistent current vanishes.
For this reason only the non-zero Fourier components
of the Coulomb-potential were retained in Ref.\cite{Kopietz93}.
At long wavelengths these
can be expressed in terms of generalized capacitances 
$C_k$, describing long-wavelength charge fluctuations 
which do not change the total electron number.
On the other hand, in a grand canonical ensemble 
the particle number $N ( \mu , \varphi )$ is a very complicated
flux-dependent random variable, which has to be determined
self-consistently by solving Eq.\ref{eq:Nselfcon}. 
By making the approximation \ref{eq:Hselfcon}
on the right-hand side of Eq.\ref{eq:AGI}, 
we can estimate the interaction contribution to the canonical
persistent current in a way that is very well suited for
numerical calculations.
For the simplified capacitance model
Eq.\ref{eq:Nselfcon} reads 
 \begin{equation}
 N ( \mu ,\varphi ) = 
 \sum_{\alpha} \Theta ( \tilde{\mu} - \epsilon_\alpha ( \varphi )  - 
 \frac{e^2  N ( \mu , \varphi) }{C_0} )
 \; \; \; , \; \; \; \tilde{\mu} = \mu + \frac{e^2 N_{+}}{C_0}
 \label{eq:Nselfcon1}
 \; \; \; ,
 \end{equation}
where the energies $\epsilon_{\alpha} ( \varphi )$ are the 
solutions of Eq.\ref{eq:schroedinger} for $U_H = 0$.
Eq.\ref{eq:Nselfcon1} can be viewed as the simplest form of self-consistent
Hartree theory. 

To appreciate the non-perturbative nature of
Eq.\ref{eq:Nselfcon1}, let us  calculate the average 
compressibility $\kappa ( \mu , \varphi )$ defined in
Eq.\ref{eq:kappadef}.
Therefore we simply differentiate both sides of Eq.\ref{eq:Nselfcon1} with respect to 
$\mu$ (taking into account that
 ${\partial  N_{+} }/ { \partial \mu } = 0$, because 
the positive background charge is fixed), solve
for $\partial  N / \partial \mu $, and finally average. This  yields
 \begin{equation}
 \kappa (\mu , \varphi ) =
  \overline{ 
  \rho ( \mu , \varphi ) \left[ 
 { 1 + \frac{e^2}{C_0} \rho ( \mu , \varphi ) }
 \right]^{-1}} 
 \label{eq:Nmu1}
 \; \; \; ,
 \end{equation}
where the self-consistent density of states is given by
 \begin{equation}
 \rho ( \epsilon , \varphi ) = 
  \sum_{\alpha} \delta ( \epsilon - \epsilon_\alpha ( \varphi ) - 
 \frac{e^2 ( N ( \mu , \varphi) - N_{+})}{C_0} )
 \label{eq:rhoe}
 \; \; \; .
 \end{equation}
Ignoring random fluctuations of the density of states, we may
factorize the average,
 \begin{equation}
  \overline{ 
  \rho ( \mu , \varphi ) \left[ 
 { 1 + \frac{e^2}{C_0} \rho ( \mu , \varphi ) }
 \right]^{-1} } 
 \rightarrow
  \overline{\rho ( \mu , \varphi ) }
  \left[ 
  1 + \frac{e^2}{C_0} \overline{ \rho ( \mu , \varphi  )} \right]^{-1}
 \label{eq:fac}
 \; \; \; .
 \end{equation}
In the diffusive regime
it is also reasonable to neglect interaction contributions to the
average density of states\cite{Altshuler92}, so that
 \begin{equation}
 \overline{ \rho ( \mu , \varphi ) } \rightarrow 
 \overline{ \rho_0 ( \mu , \varphi ) }  
 \equiv
 \overline{ 
  \sum_{\alpha} \delta ( \mu - \epsilon_\alpha ( \varphi )  ) } = \frac{1}{\Delta }
  \; \; \; ,
  \label{eq:rho0def}
 \end{equation}
where $\Delta$ is the average level spacing at the Fermi energy
in the absence of interactions.
Hence we obtain
 \begin{equation}
  \kappa ( \mu , \varphi ) = Z_{\rm RPA}  \Delta^{-1}
 \; \; \; , \; \; \;
 Z_{\rm RPA} = \frac{1}{1 + \frac{e^2}{C_0 \Delta } }
 \; \; \; . 
 \label{eq:kapparen}
 \end{equation}
Note that $ \Delta^{-1}$ is the average compressibility
without interactions.
Because ${e^2}/({C_0 \Delta }) \gg 1$
in the experimentally relevant parameter regime, 
the RPA renormalization factor $Z_{\rm RPA}$ is small compared with unity.
Hence, the interactions 
lead to a drastic reduction of the average compressibility. 
A similar result has also been obtained by Vignale
within a density functional approach\cite{Vignale94a}.
A strong reduction of the compressibility has also been observed
in the exact diagonalization study by Berkovits and Avishai\cite{Berkovits96}.
The origin for this effect is
intuitively obvious,  because
physically $\kappa^{-1}$ is the energy
associated with the addition of an electron to the system.
In the absence of interactions this is simply the average level spacing,
but with Coulomb interactions one has to pay the
charging energy ${e^2}/{C_0}$ for adding an electron. 
Hence, the inverse compressibility is pushed from 
$\Delta$ to the much larger energy ${e^2}/{C_0}$.
We would like to stress that in the conventional diagrammatic
approach to the many-body problem for
infinite translationally invariant systems the 
Hartree diagram responsible for this renormalization
is usually ignored, because
it is assumed to be canceled exactly by the
positive background charge\cite{Nozieres64}.
Substituting Eq.\ref{eq:kapparen} into Eq.\ref{eq:AGI}, we obtain
 \begin{eqnarray}
 \overline{I ( N , \varphi )} 
 -
 \overline{I ( \mu^{\ast} , \varphi )}
 & \approx  &
    \frac{- e}{ h } \frac{\Delta}{2 Z_{\rm RPA} }
 \frac{ \partial}{\partial \varphi }
  \Sigma^{2} ( \mu^{\ast} , \varphi )
 \nonumber
 \\
 & \approx  &  \frac{- e}{ h } \frac{e^2}{2 C_0}
 \frac{ \partial}{\partial \varphi }
  \Sigma^{2} ( \mu^{\ast} , \varphi )
 \; \; \; ,
 \label{eq:mapping2}
 \end{eqnarray}
where the second line is valid for $e^2 / C_0 \gg \Delta$.
The crucial question is now whether the flux-dependent part
of the variance
  $\Sigma^{2} ( \mu^{\ast} , \varphi )$
is reduced by the interactions or not. Note that naive application of the standard 
RPA\cite{Vignale94,Vignale94a}
 leads to the
replacement
$  \Sigma^{2} ( \mu^{\ast} , \varphi )
 \rightarrow Z^2_{\rm RPA} 
  \Sigma^{2}_0 ( \mu^{\ast} , \varphi )$,
where 
  $\Sigma^{2}_0 ( \mu^{\ast} , \varphi )$ is the variance
in the absence of interactions (see Eq.\ref{eq:Sigmascreen} below).
However, as already mentioned, the {\it{flux-dependent}} part of
  $\Sigma^{2} ( \mu^{\ast} , \varphi )$ 
is much smaller than the flux-independent part, and it
is not at all clear whether it is
screened in exactly the same way as
the flux-independent part.
Because even without disorder it is very difficult to calculate
corrections to the RPA in a systematic way\cite{Mattuck76}, we shall
in this work study this problem numerically.

\section{Numerical results}
\label{sec:num}

We have calculated the eigenvalues
$E_{\alpha} ( \varphi ) = \epsilon_{\alpha} ( \varphi ) + 
\frac{ e^2}{C_0} ( N ( \mu , \varphi ) - N_{+} )$
of Eq.\ref{eq:schroedinger} numerically for fixed realizations of the random
potential $U ( {\bf{r}} )$ on finite lattices.
The eigenvalues $\epsilon_{\alpha} ( \varphi )$ without
the Hartree term were calculated with the Lanczos method\cite{Cullum85,Holm95}.
To obtain the energy shift due to the Hartree potential, 
the self-consistency equation 
\ref{eq:Nselfcon1} for the particle number for a given realization
of the random potential was then solved via a simple
Newton procedure. 
It turns out that at zero temperature this equation does not 
always have a unique solution, because by definition $N$ must be an integer.
This problem is easily avoided by working at finite temperature $T$,
and then extrapolating for $T \rightarrow 0$.
Because for $T > 0$ the step function in Eq.\ref{eq:Nselfcon1} is smoothed out
into the Fermi function, there exists always a unique solution of
Eq.\ref{eq:Nselfcon1}. This is easy to see from the fact that
the left-hand side of Eq.\ref{eq:Nselfcon1} is a monotonically increasing
function of the particle number, while the right-hand side is monotonically decreasing.

\begin{figure}
\vspace{1cm}
\hspace{1cm}
\epsfysize10cm 
\epsfbox{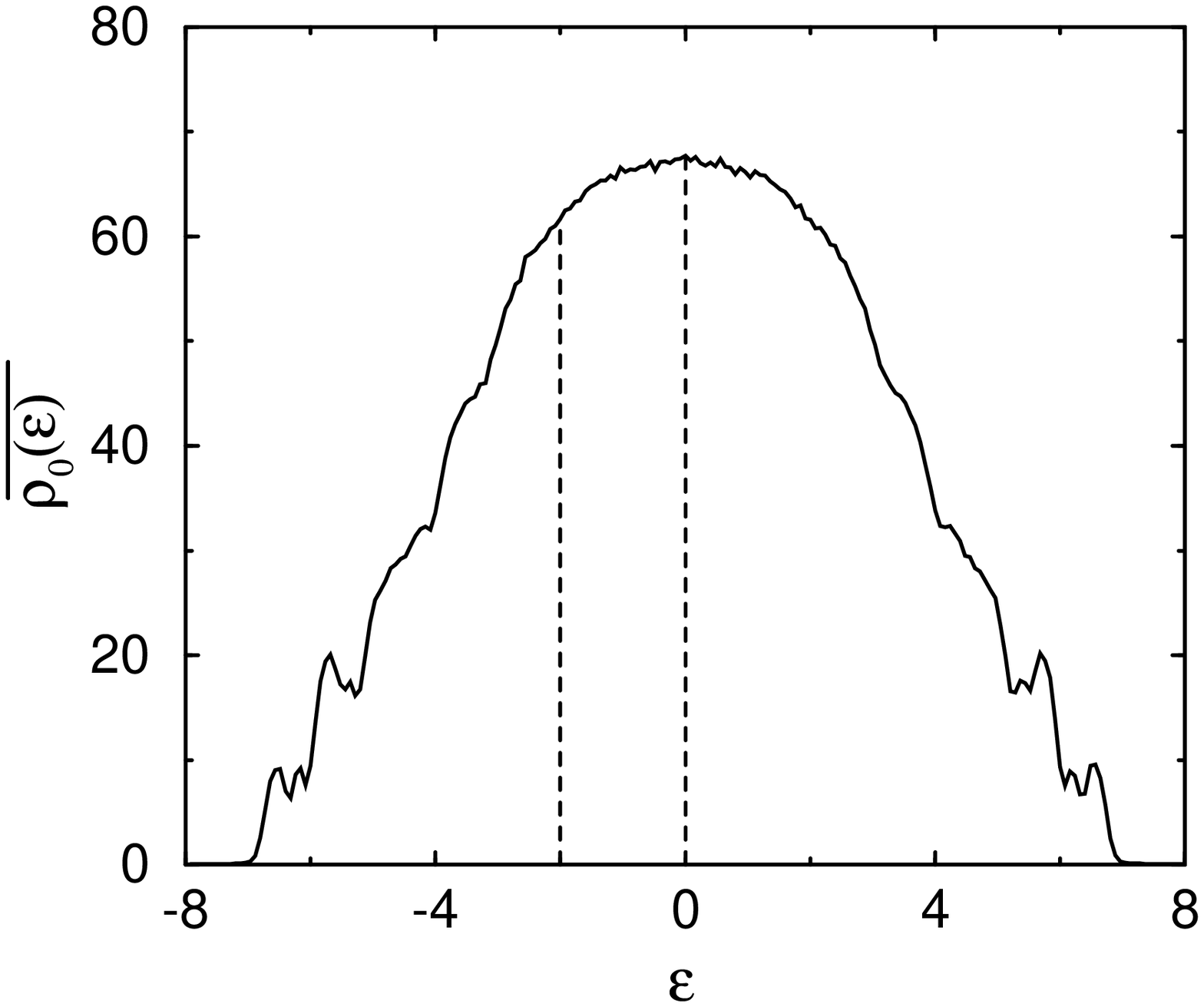}
%\vspace{1cm}
\caption{Non-interacting average density of states 
of a $20 \times 5 \times 5$ system
with $w = 4$.
For our numerical calculations we discarded all energies
outside the interval $[\mu - E_0 , \mu ]$ with $\mu=0 , E_0=2$ (see the
dashed lines) in order
to obtain 
a model with approximately constant density of states.
}
\label{fig:dos}
\end{figure}

Most theoretical calculations in disordered systems 
are based on the assumption that in the energy window of interest the
energy-dependence of the average density of states can be ignored.
On the other hand, in our finite system the average density of states 
is non-zero only in a finite interval, and exhibits a broad
maximum at zero energy, as shown in Fig.\ref{fig:dos}.
As we have recently pointed out\cite{Voelker96}, 
for non-interacting  electrons the universal 
weak localization effects related to the Cooperon pole
are completely washed out if the energy-dependence of the
average density of states becomes significant in the energy interval 
under consideration. 
Hence, 
in order to compare our numerical calculations with
theoretical predictions for models with constant
density of states, we shall retain only the central part
of the spectrum, and discard all energies outside an interval
$[ \mu - E_0 , \mu]$. 
From Fig.\ref{fig:dos} it is obvious that for 
our set of parameters $E_0 = 2$ is a good choice.
A similar procedure has also been adopted in Ref.\cite{Braun95}.
Note that in this way we formally replace the
density of states of our original tight binding model
by the density of states with sharp edges shown in Fig.\ref{fig:dos}. 
The total particle number is still  defined
as in Eq.\ref{eq:Nselfcon1}, except that now the $\alpha$-sum is over
energies satisfying $\epsilon_{\alpha} \in [ \mu - E_0 , \mu ]$.
The great advantage of this construction is that the average density of states
is practically constant in the entire regime where it is non-zero,
so that non-universal effects related to the energy-dependence of the density of states
should be clearly visible at the band edges.

\begin{figure}
\vspace{1cm}
\hspace{1cm}
\epsfysize10cm 
\epsfbox{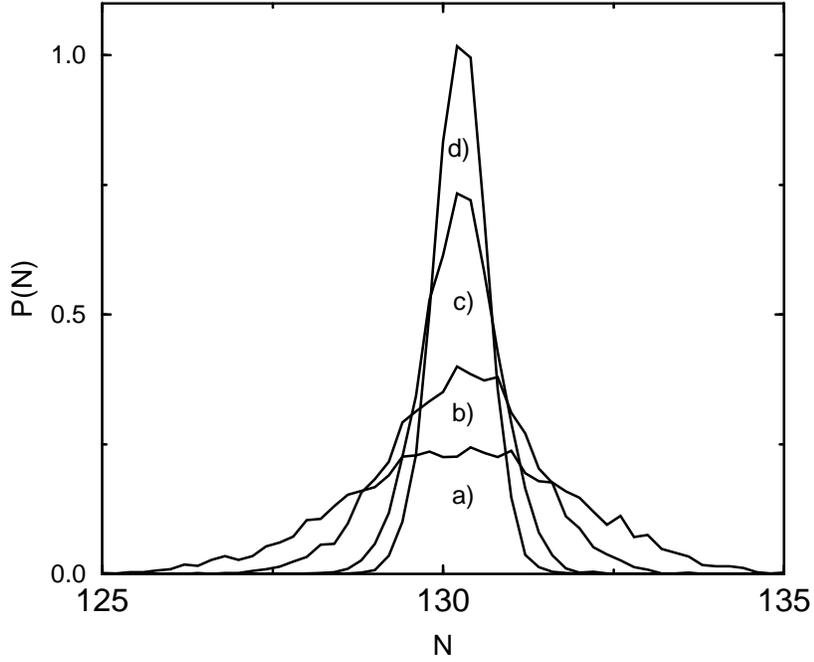}
%\vspace{1cm}
\caption{
Probability distribution $P ( N )$ of the particle number 
in the simplified capacitance model 
on a $20 \times 5 \times 5$ lattice
for different values of $e^{2} / ( C_0 \Delta )$,
with $\Delta^{-1}=\overline{\rho_0 (\mu)}=67$.  
(a) $e^2 / ( C_0 \Delta ) =  0$;
(b) $e^2 / ( C_0 \Delta ) =  \Delta^{-1}/100 =  0.67$;
(c) $e^2 / ( C_0 \Delta ) =  3\Delta^{-1}/100 = 2.01 $;
(d) $e^2 / ( C_0 \Delta ) =  \Delta^{-1}/20 = 3.35 $.
Here and in all 
subsequent figures we have chosen
$w = 4$, $\mu = 0$ and $N_{+} = 130.2$. 
In this case $N_{+}$ agrees with the average particle number, so that
on average the ensemble of rings is not charged.
}
\label{fig:probab}
\end{figure}

\begin{figure}
\vspace{1cm}
\hspace{1cm}
\epsfysize10cm 
\epsfbox{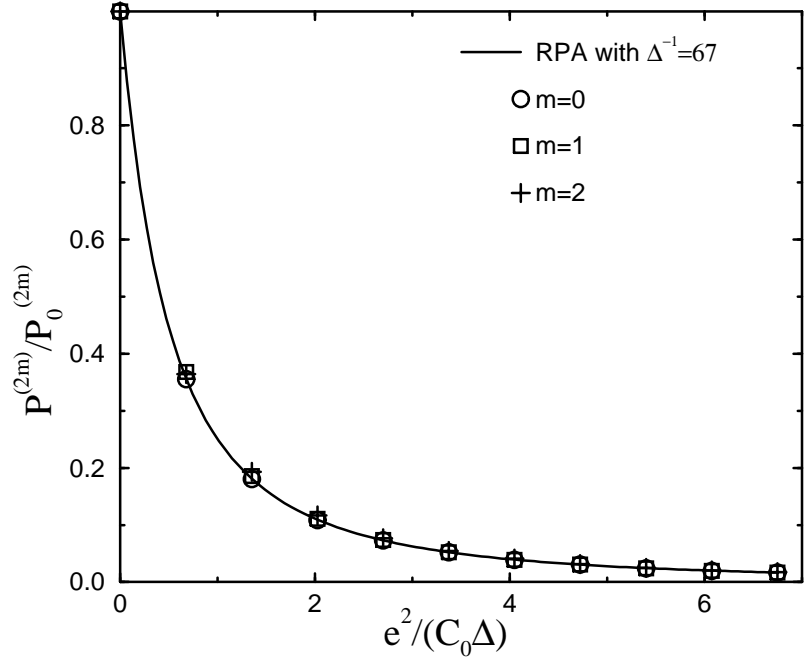}
%\vspace{1cm}
\caption{
Ratio of the even Fourier components
$P^{(2m)} / P^{(2m)}_0$ (see Eq.\ref{eq:Fourierexp})
of the particle number variance for different values of $m$
as function of $e^2 / ( C_0 \Delta )$.
The solid line is the RPA prediction $Z_{\rm RPA}^2$.
}
\label{fig:RPAtest}
\end{figure}

The drastic effect of the Coulomb interaction 
on the statistics of the number of electrons in the system
is clearly seen in Fig.\ref{fig:probab}, were we 
plot the probability distribution $P ( N )$ of the particle number
for different values of $e^2 / C_0$.
The narrowing of the distribution for increasing
interaction parameter can be viewed  as a screening effect.
The usual RPA prediction 
for the second moment $\Sigma^{2} ( \mu , \varphi ) = \overline{ ( N ( \mu , \varphi)  - N_{+} )^2}$ 
of the distribution
$P ( N )$ around $N_{+}$ is easily obtained from Eq.\ref{eq:Nselfcon1} by
expanding the right-hand side to {\it{first order}} in
$ N - N_{+}$,
 \begin{equation}
 N - N_{+} = N_0
 - N_{+} 
 - \frac{{e}^2}{C_0} \rho_0 ( \mu , \varphi )
 ( N - N_+ )
 + O( ( N - N_{+} )^2)
 \label{eq:RPAexpand}
 \; \; \; ,
 \end{equation}
where $N_0 =
 \sum_{\alpha} \Theta ( {\mu} - \epsilon_\alpha  )$ is the particle number
in the absence of interactions, and $\rho_0 ( \mu , \varphi )$
is the non-interacting density of states, see Eq.\ref{eq:rho0def}.
Solving Eq.\ref{eq:RPAexpand} for $N -  N_{+}$, squaring, averaging, and
ignoring random fluctuations of the density of states (see 
Eq.\ref{eq:fac}), we obtain
for the second moment
 \begin{equation}
 \Sigma^2 ( \mu , \varphi ) = Z_{\rm RPA}^2 \Sigma_0^2 ( \mu , \varphi )
 \; \; \; ,
 \label{eq:Sigmascreen}
 \end{equation}
where $Z_{\rm RPA}$ is given in Eq.\ref{eq:kapparen},
and $\Sigma_0^{2} ( \mu , \varphi ) = \overline{ (N_0 ( \mu , \varphi ) - N_{+} )^2}$.
The obvious question is now whether 
Eq.\ref{eq:Sigmascreen} is accurate or not. Note that
this expression is based on the
expansion \ref{eq:RPAexpand}, which is a priori uncontrolled, because
it is not clear whether the effect of the higher order terms
that have been ignored
is important or not. Furthermore, it is by no means clear whether  
the {\it{flux-dependent}} part of $\Sigma^2 ( \mu , \varphi )$ is
modified in exactly the same way by the interaction as the dominant
flux-independent part. To investigate this point, let us expand 
$\Sigma^2 ( \mu , \varphi )$ in a Fourier series. 
Because
$N ( \mu , \varphi )$ is an even  periodic function
of $\varphi$ with a fundamental period of unity, we may expand
 \begin{equation}
 \Sigma^2 ( \mu , \varphi ) 
 = \frac{P^{(0)}}{2} + \sum_{n=1}^{\infty} P^{(n)} \cos ( 2 \pi n \varphi )
 \; \; \; .
 \label{eq:Fourierexp}
 \end{equation}
Denoting by $P_0^{(n)}$ the corresponding Fourier components
in the absence of interactions, the RPA predicts that
$P^{(n)} / P_0^{(n)} = Z^2_{\rm RPA}$ for all $n$.
As shown in Fig.\ref{fig:RPAtest}, for the zeroth and the
first two even Fourier components $(n = 2m = 0,2,4)$
this prediction is in excellent quantitative
agreement with our numerical
self-consistent solution of Eq.\ref{eq:Nselfcon1}. The odd components
$( n = 2m+1)$
are numerically found to be much smaller than
the even ones\cite{Voelker96}, but are screened in a similar fashion.
We conclude that for our simplified capacitance model 
the RPA is an excellent approximation,
so that the interaction actually reduces the
current in Eq.\ref{eq:mapping2}
by a factor of $Z_{\rm RPA} \ll 1$. 
In this case the 
difference between the canonical and the grand canonical current
is completely negligible compared with the leading contribution to the 
average grand canonical
current $\overline{I ( \mu^{\ast} , \varphi )}$ that has been 
considered by Ambegaokar and Eckern\cite{Ambegaokar90}.

However, the screening behavior is completely changed if we 
exclude the energies in the vicinity of the band edges.
Such a procedure can be motivated physically as follows:
It is well-known that in an interacting many-body system  
propagating quasi-particles  exist only for energies sufficiently close to the Fermi energy.
Moreover,  the
flux-dependence of physical observables should
be due to quasi-particles that can
coherently propagate around the ring,
and in this way probe the sensitivity to twists in the boundary conditions.
Hence,
we expect that in a realistic interacting many-body system
$\partial \Sigma^2 ( \mu , \varphi ) / \partial \varphi $
is essentially determined by weakly damped states with energies in a small
interval of width $E_{\rm in} $ below the Fermi energy,
Here
$E_{\rm in}$ is some inelastic cutoff energy that is small compared with
the Fermi energy.
In other words,
inelastic processes (which are completely ignored in the Hartree and the Hartree-Fock 
approximation) will naturally restrict the states contributing
to the {\it{flux-dependent}} part of $\Sigma^{2} ( E , \varphi )$
to those with energies in the interval $[ \mu - E_{\rm in} , \mu ]$.
The theory developed in Refs.\cite{Kopietz93,Kopietz94a}
is based on the existence of such a cutoff energy.

To study the contributions from the part of the spectrum
close to the Fermi energy to $\Sigma^2 ( \mu , \varphi )$,
let us therefore introduce the auxiliary quantity
 \begin{equation}
\sigma^{2} ( E , \mu , \varphi ) 
 = \int_{\mu -E}^{\mu} d \epsilon
  \int_{\mu -E}^{\mu} d \epsilon^{\prime} 
  K_2 ( \epsilon , \epsilon^{\prime} , \varphi )
 \label{eq:NKdef}
 \; \;\; ,
 \end{equation}
 \begin{equation}
  K_2 ( \epsilon , \epsilon^{\prime} , \varphi )
  = \overline{
 \rho ( \epsilon , \varphi )  
 \rho ( \epsilon^{\prime} , \varphi )  }
 - \overline{ \rho ( \epsilon , \varphi )} \; \;
 \overline{ \rho ( \epsilon^{\prime} , \varphi )} 
 \; \; \; ,
 \label{eq:Kdef}
 \end{equation}
where the 
self-consistent density of states $\rho ( \epsilon , \varphi )$ 
is given in Eq.\ref{eq:rhoe}.  Note that 
$\sigma^{2} ( E , \mu , \varphi ) $ is the variance of the number of energy levels
in an interval of width $E$ below the Fermi energy,
and that by construction
$\Sigma^{2} ( \mu , \varphi ) 
= \lim_{E \rightarrow \infty} \sigma^{2} ( E, \mu , \varphi ) $.
In fact, given  our truncated density of states shown
in Fig.\ref{fig:dos},
we have 
$\Sigma^{2} ( \mu , \varphi ) =
\sigma^{2} ( E, \mu , \varphi ) $ for all $E > E_0$.
It is again convenient to expand 
$\sigma^{2} ( E , \mu , \varphi ) $ in a Fourier series,
 \begin{equation}
 \sigma^2 ( E, \mu , \varphi ) 
 = \frac{p^{(0)} ( E )}{2} + \sum_{n=1}^{\infty} p^{(n)} (E) \cos ( 2 \pi n \varphi )
 \label{eq:FourierSigma}
 \; \; \; .
 \label{eq:pFourierexp}
 \end{equation}
In the absence of interactions and for energies $E$  small compared with
$\hbar / \tau$ (where $\tau$ is the elastic lifetime)
an approximate expression for the Fourier coefficients $p^{(n)}_{0} ( E )$ 
(where the subscript $0$ indicates that interactions are neglected) can
be obtained from the Feynman diagrams with two Cooperons and two
Diffusons given by Altshuler and Shklovskii\cite{Altshuler86}.
In this approximation one obtains $p_{0}^{(2m+1)} ( E ) = 0$, and 
the even Fourier components are given by\cite{Voelker96}
 \begin{eqnarray}
 p^{(2m)}_{0} ( {E} )  & = & \frac{2}{\pi^2 m}
 \left\{  
 \exp \left[ -m \tilde{\Gamma}^{1/2} \right]
 - 
  \exp \left[
 -  \frac{m}{\sqrt{2}}
 \left( 
  \sqrt{ \tilde{E}^2 + \tilde{\Gamma}^2 } + 
  \tilde{\Gamma}
  \right)^{1/2} 
  \right]
 \nonumber
 \right.
 \\
 & & 
 \left.
 \hspace{4cm}
 \times \cos
 \left[ \frac{m}{\sqrt{2}}
 \left( 
   \sqrt{\tilde{E}^2 + \tilde{\Gamma}^2 } - \tilde{\Gamma} \right)^{1/2}  
  \right]
 \right\}
 \; \; \; .
 \label{eq:P0mE}
 \end{eqnarray}
The flux average is 
 \begin{eqnarray}
 \frac{p_0^{(0)} ( {E} )}{2} & = & \frac{2}{ \pi^2}
 \left[ 
 \frac{1}{\sqrt{2}}
 \left( 
  \sqrt{ \tilde{E}^2 + \tilde{\Gamma} } + \tilde{\Gamma} \right)^{1/2}  
    -  {\tilde{\Gamma}}^{1/2} \right]
 \nonumber
 \\
 & + & \sum_{m=1}^{\infty} p_0^{(2m)} ( \tilde{E})
 \label{eq:P00E}
 \; \; \; .
 \end{eqnarray}
Here $\tilde{E} = E / E_c$, and $\tilde{\Gamma} = \Gamma / E_c$,
where $E_c$ is the Thouless energy and $\Gamma$ is some 
cutoff energy that has been introduced into the non-interacting theory
{\it{by hand}}\cite{Schmid91}.
For non-interacting electrons on a quasi-one dimensional
ring (where diffusion is only possible along the circumference)
we have recently confirmed
Eqs.\ref{eq:P0mE} and \ref{eq:P00E} numerically\cite{Voelker96}.
Here we would like to investigate whether in the presence
of long-range Coulomb interactions these expressions are strongly
reduced by screening effects.  
Note that according to Altland and Gefen\cite{Altland94}
there should be no qualitative differences between the screening
of fluctuations of the
{\it{spectral}} density and the {\it{total}} density; 
if this would be correct,
then the variance $\sigma^2 ( E , \mu , \varphi )$ of the number of 
energy levels
in an interval $E$ below the Fermi energy should be screened in precisely
the same way as the variance
$\Sigma^2 ( \mu , \varphi )$ of the total particle
number (which is strongly affected by screening, see Fig.\ref{fig:RPAtest}).
We are now in the position to settle this controversy
numerically within the simplified capacitance model. 
We would like to emphasize that, in spite of its simplicity, this model
contains the physics of screening.

\begin{figure}
\vspace{1cm}
\hspace{1cm}
\epsfysize10cm 
\epsfbox{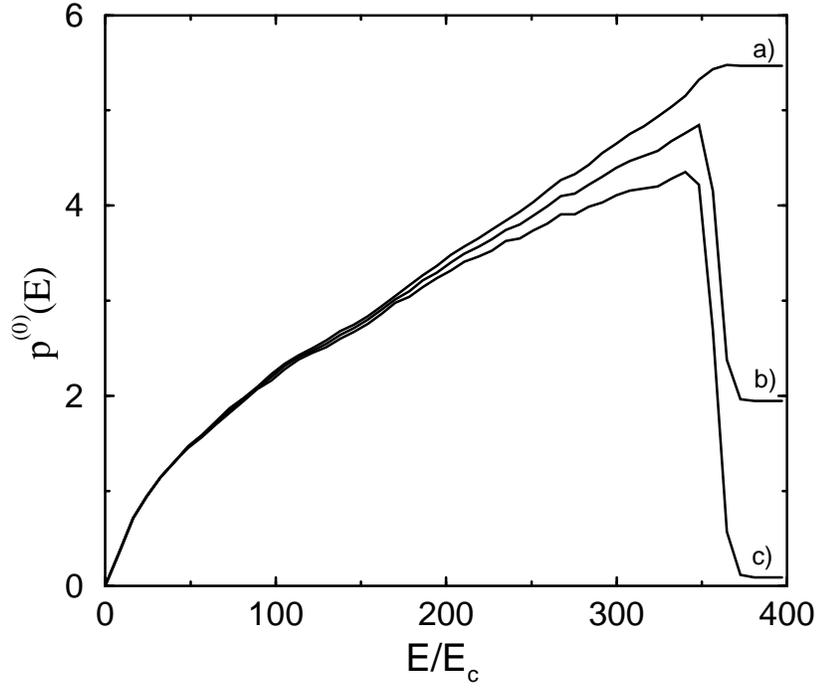}
%\vspace{1cm}
\caption{
Zeroth Fourier component $p^{(0)} ( E ) $ of the variance
$\sigma^{2} ( E , \mu , \varphi )$ as function of $E/E_c$
with $E_c=1/180$ and $\Delta^{-1}=67$ for
different 
values of $e^2 / (C_0 \Delta )$.
(a) $e^2 / ( C_0 \Delta ) =  0$;
(b) $e^2 / ( C_0 \Delta )  = 0.67$;
(c) $e^2 / ( C_0 \Delta ) = 6.7$.
The band edge is at $E_0/E_c=360$.}
\label{fig:screenE0}
\end{figure}

\begin{figure}
\vspace{1cm}
\hspace{1cm}
\epsfysize10cm 
\epsfbox{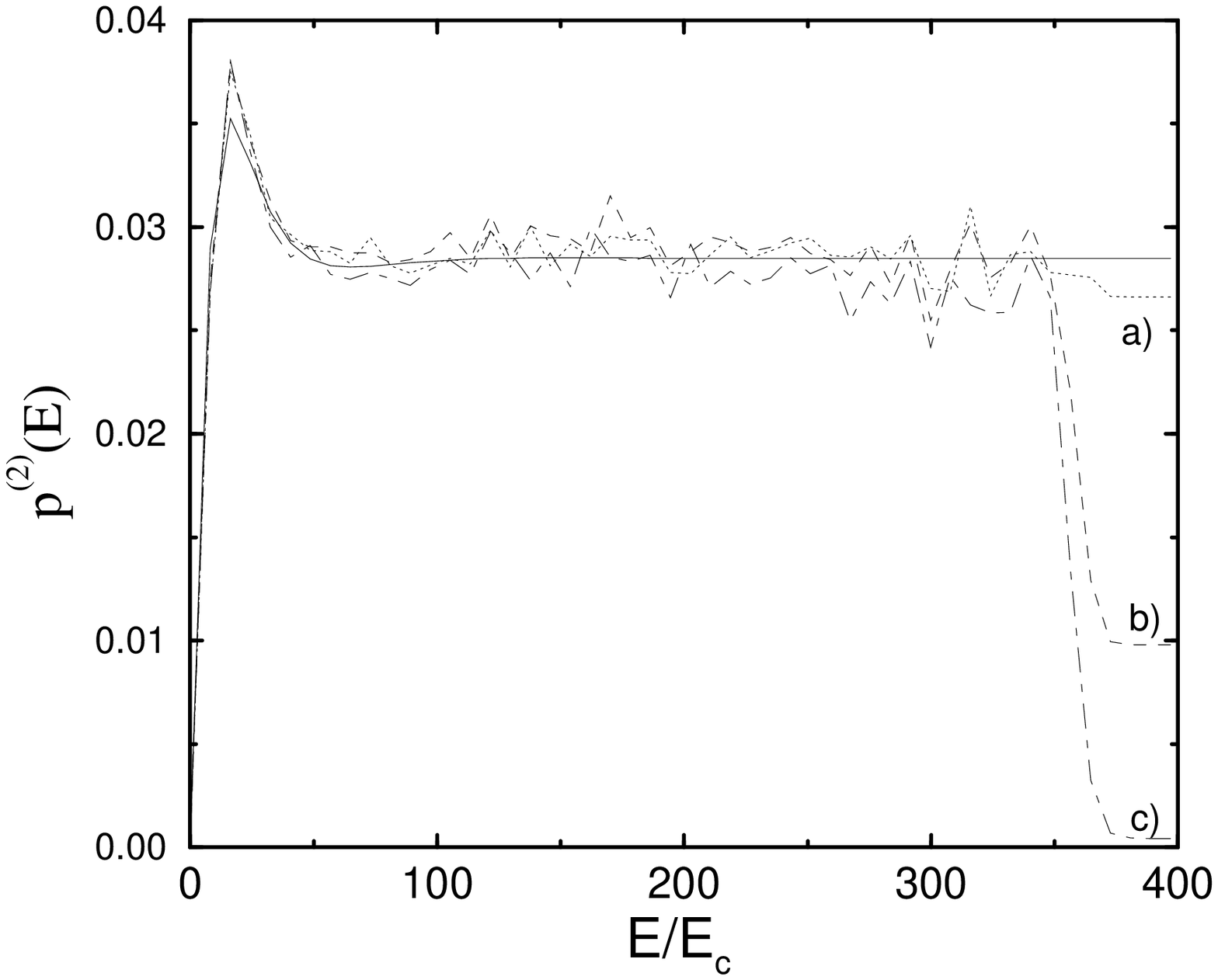}
%\vspace{1cm}
\caption{
Second Fourier component $p^{(2)} (  E ) $ of the variance
$\sigma^{2} ( E , \mu , \varphi )$ as function of $E/E_c$ for
different 
values of $e^2 / (C_0 \Delta )$ with $\Delta^{-1}=67$.
(a) $e^2 / ( C_0 \Delta ) =  0$;
(b) $e^2 / ( C_0 \Delta ) =  0.67 $;
(c) $e^2 / ( C_0 \Delta ) = 6.7 $.
The solid line is the Altshuler-Shklovskii
prediction \ref{eq:P0mE} with $E_c = 1/180$ and ${\tilde\Gamma} = 3.85$.
}
\label{fig:screenEphi}
\end{figure}

In Fig.\ref{fig:screenE0} 
we show our numerical results for the zeroth Fourier component 
$p^{(0)} ( E ) $ 
as function of $E$ for different values of
$e^{2} / ( C_0 \Delta )$.
Corresponding results for the first even Fourier component
$p^{(2)} ( E ) $ 
are shown in Fig.\ref{fig:screenEphi}.
Evidently, in the entire energy interval where the average density of states
is approximately constant, the flux average of $\sigma^{2} ( E , \mu , \varphi )$ and the
dominant flux-dependent contribution are only weakly affected by
the Coulomb interaction. Moreover, in Fig.\ref{fig:screenEphi} we also
show the Altshuler-Shklovskii prediction \ref{eq:P0mE} 
for $p^{(2)} ( E )$. The agreement with the numerical results
for the simplified  capacitance  model clearly demonstrates that for small
energies it is indeed allowed to ignore screening corrections to
$\sigma^{2} ( E , \mu , \varphi )$. Thus, our numerical results support the
arguments put forward by one of us in 
Refs.\cite{Kopietz93,Kopietz94a,Kopietz94b,Kopietz94c}, and strongly disagree with
the statement of Altland and Gefen\cite{Altland94} that the fluctuations of the
spectral density is screened in the  same way as the total density.
From Figs.\ref{fig:screenE0} and \ref{fig:screenEphi}
it is also obvious that the RPA screening of the fluctuation of the total particle number
is recovered as soon as the interval $E$ includes the band edge $E_0$.
The sharp drop for $p^{(0)} ( E )$ and $p^{(2)} ( E )$ for $E \approx E_0$ is
in agreement with the RPA result for the Fourier components $P^{(n)}$ of
the variance of the total particle number shown in Fig.\ref{fig:RPAtest}.
Of course, to obtain the physical persistent current, fluctuations on all energy scales
should be included, so that the simplified capacitance model does not show
any enhancement of the average persistent current due to long-range Coulomb interactions.
However, as explained in Refs.\cite{Kopietz94a,Voelker96},
we expect that in a realistic many body system the 
damping of quasi-particles which are not close to the Fermi surface will
eliminate the contribution from high-energy states
to the {\it{flux-dependent}} part of $\Sigma^{2} ( \mu , \varphi )$. In other words,
for a realistic interacting many-body system we expect
 \begin{equation}
 \frac{ \partial}{\partial \varphi }
 \Sigma^2 ( \mu , \varphi ) \approx
 \frac{ \partial}{\partial \varphi }
 \sigma^2 ( E , \mu , \varphi ) 
 \; \; \; , \; \; \; 
 \mbox{for all $E 
\raisebox{-0.5ex}{$\; \stackrel{>}{\sim} \;$} E_{\rm in}$}
\;  \; \; , 
\label{eq:Ein}
 \end{equation}
where $E_{\rm in}$ is the unknown inelastic cutoff energy mentioned above.
Note that in Ref.\cite{Altshuler91} Altshuler, Gefen, and Imry 
implicitly seem to make a similar assumption\cite{footnote1}.

\section{Conclusions}

In this work we have presented numerical results for the simplified 
capacitance model, which describes
the effect of long-range Coulomb interaction on the average canonical persistent
current. The model can be considered as the simplest form of self-consistent
Hartree theory, which still contains the non-trivial physics of screening.
We have confirmed one of the fundamental assumptions
of Ref.\cite{Kopietz93}, namely that the fluctuations of the {\it{spectral}} density
for sufficiently small energies are not screened.
Of course, within the approximations inherent in our model  
all eigen-energies (including those far away from the Fermi energy)
correspond to quasi-particles with infinite lifetime, so that the
physical persistent current is eventually screened once the energies
at the band edges are taken into account. However, in physically
more realistic models we expect that the damping of the quasi-particles
will automatically eliminate non-universal effects related to the band edges.
Given the fact that the low-energy fluctuations are not screened,
the universal current proposed in Ref.\cite{Kopietz93}
directly follows.

The problem of verifying by explicit calculation
that the damping of quasi-particles
restricts the flux-dependent part
of the spectrum to a small interval of energies close to the Fermi energy
remains open. We believe that this problem is closely related to
the rather obscure inelastic cutoff $\Gamma$, which is usually
introduced phenomenologically into the Cooperon
and Diffuson propagators of the non-interacting theory.
For a recent microscopic calculation of $\Gamma$ see Ref.\cite{Blanter96}.
More generally, the inelastic cutoff
should be momentum- and frequency-dependent,
$\Gamma ( {\bf{q}} , \omega )$. It is tempting to speculate that the energy scale 
where $\Gamma ( 0 , \omega )$  begins to 
deviate significantly from its zero-energy limit $\Gamma ( 0 , 0 )$
is proportional to the cutoff energy $E_{\rm in}$
mentioned above (see Eq.\ref{eq:Ein}).

\section*{Acknowledgements}
\vspace{0.2cm}
We are grateful to J. A. Holm for providing us with
his {\it{Master-Slave Implementation for Parallel Virtual Machines}}\cite{Holm95},
which opened the way for an efficient numerical simulation
on a work-station cluster.
We would also like to thank K. Sch\"{o}nhammer for discussions.
This work was financially supported by the Deutsche Forschungsgemeinschaft
(SFB 345).

%
%              R E F E R E N C E S

\begin{thebibliography}{99}
\bibitem{footnote1}
In Ref.\cite{Altshuler91} Altshuler, Gefen and Imry claim 
(see the text before their Eq.9) that
for the calculation of
$\Sigma^2 ( \mu , \varphi ) - \Sigma^2 ( \mu , 0 )$
it is sufficient to consider the correlator
$K_2  ( \epsilon , \epsilon^{\prime} , \varphi )$ only for energies
in the vicinity of the Fermi energy.
For non-interacting electrons this is certainly not the case\cite{Voelker96},
so that one has to invoke inelastic processes in order
to justify this assumption.
\end{thebibliography}
\end{document}